\documentclass[prd,reprint,twocolumn,aps,showkeys,fleqn,
superscriptaddress,groupedaddress,unsortedaddress,runinaddress,
frontmatterverbose,preprintnumbers,nofootinbib,nobibnotes,bibnotes,
prstab,prstper,floatfix,amsmath,amssymb]{revtex4-2}

\usepackage[T1]{fontenc}
\usepackage{graphicx}
\usepackage{float}
\usepackage{amsmath}
\usepackage{multirow}
\usepackage{dcolumn}
\usepackage{bm}
\usepackage{afterpage}
\usepackage{placeins}
\usepackage{booktabs}
\usepackage[dvipsnames]{xcolor}

\usepackage[hidelinks]{hyperref}
\usepackage{cleveref}
\hypersetup{colorlinks=true,linkcolor=blue,urlcolor=blue,citecolor=blue}

\usepackage[left,mathlines]{lineno}
\begin{document}
\title{Energy calibration of LHAASO-KM2A using the cosmic ray Moon shadow}

\author{Zhen Cao$^{1,2,3}$,
F. Aharonian$^{3,4,5,6}$,
Y.X. Bai$^{1,3}$,
Y.W. Bao$^{7}$,
D. Bastieri$^{8}$,
X.J. Bi$^{1,2,3}$,
Y.J. Bi$^{1,3}$,
W. Bian$^{7}$,
A.V. Bukevich$^{9}$,
C.M. Cai$^{10}$
% ...
}

\affiliation{$^{1}$ Key Laboratory of Particle Astrophysics \& Experimental Physics Division \& Computing Center, Institute of High Energy Physics, Chinese Academy of Sciences, 100049 Beijing, China\\
$^{2}$ University of Chinese Academy of Sciences, 100049 Beijing, China\\
$^{3}$ TIANFU Cosmic Ray Research Center, Chengdu, Sichuan,  China\\
$^{4}$ University of Science and Technology of China, 230026 Hefei, Anhui, China\\
$^{5}$ Yerevan State University, 1 Alek Manukyan Street, Yerevan 0025, Armenia\\
$^{6}$ Max-Planck-Institut for Nuclear Physics, P.O. Box 103980, 69029  Heidelberg, Germany\\
$^{7}$ Tsung-Dao Lee Institute \& School of Physics and Astronomy, Shanghai Jiao Tong University, 200240 Shanghai, China\\
$^{8}$ Center for Astrophysics, Guangzhou University, 510006 Guangzhou, Guangdong, China\\
$^{9}$ Institute for Nuclear Research of Russian Academy of Sciences, 117312 Moscow, Russia\\
$^{10}$ School of Physical Science and Technology \&  School of Information Science and Technology, Southwest Jiaotong University, 610031 Chengdu, Sichuan, China\\
% ...
}

\date{\today}

\begin{abstract}
We present a precise measurement of the westward rigidity-dependent shift of the Moon’s shadow using three and a half years of cosmic ray data collected by the kilometer square array (KM2A) of the Large High Altitude Air Shower Observatory (LHAASO) experiment. These measurements enable us to calibrate the detector energy response in the range of 20 to 260 TeV, with results showing excellent agreement with the energy response derived from Monte Carlo (MC) simulations of the KM2A detector. We also measure a best-fit parameter $\epsilon=0.015\pm0.08$, corresponding to a 95\% CI of $[-14\%,\, +17\%]$ for the energy scale estimation.
This result establishes the exceptional accuracy of the KM2A-MC in simulating the detector’s response within this energy range.

\end{abstract}


\maketitle
 
\begingroup
\renewcommand\thefootnote{\fnsymbol{footnote}} 
\maketitle
\footnotetext[1]{\label{fn:collab}Correspondence:
\href{mailto:aliraza@ihep.ac.cn}{\nolinkurl{aliraza@ihep.ac.cn}},
\href{mailto:hhh@ihep.ac.cn}{\nolinkurl{hhh@ihep.ac.cn}},
\href{mailto:faisal.chep@pu.edu.pk}{\nolinkurl{faisal.chep@pu.edu.pk}}}
\endgroup
\section{Introduction} \label{sec:intro}

Accurate energy calibration is crucial to the scientific reliability of extensive air shower experiments, as it directly influences the reconstruction of the cosmic ray energy spectrum and the interpretation of its key features, such as the \textit{knee} and \textit{ankle}. Even modest energy scale uncertainties can shift spectral break positions, leading to incorrect conclusions about cosmic ray acceleration limits or the galactic-to-extragalactic transition. Moreover, energy calibration affects composition analyses, since different primary nuclei exhibit distinct rigidity-dependent cutoffs and air-shower development profiles. Inconsistent energy scales across experiments such as HAWC \cite{Hawcuncertainity}, AS$\gamma$ \cite{Tibetasgamma}, and LHAASO \cite{zhyprl} have historically contributed to discrepancies in the reported spectrum, underscoring the importance of robust, cross-validated calibration strategies.

Historically, these calibration issues have had observable consequences. Early balloon-borne experiments such as JACEE \cite{JACEEOJ}, which used X-ray film and emulsion plate techniques, faced significant energy calibration uncertainties and had low statistics at the highest energies. Reports of spectral breaks or changes in the spectral index from these measurements were not confirmed by later space-borne missions. More recent experiments, including CREAM \cite{CREAM3}, DAMPE \cite{DAMPE2019}, and others, which benefit from superior exposure and precise accelerator-based calibration, consistently observe single or smoothly broken power laws for protons and iron up to $\sim$100–200 TeV, with no sharp spectral features. This highlights that some spectral index changes seen in earlier data may have been artifacts of poor energy calibration.

In contrast, ground-based arrays relying on the extensive air shower (EAS) technique and Monte Carlo (MC) simulations with hadronic interaction models have reported an all-particle knee with varying sharpness and positions. For example, AS$\gamma$ \cite{Tibetasgamma} reports a smoother knee near $\sim$4 PeV, yet suffers uncertainties of $\sim$10\% from hadronic interaction models and $\sim$20\% from composition assumptions, as summarized in Table-5 of Ref.~\cite{Tibetasgamma}.
A recent analysis using LHAASO all-particle spectrum and $\langle ln A\rangle$ reports a proton knee at  $(3.2 \pm 0.2)$ PeV and an iron-associated ankle at $(9.7 \pm 0.2)$ PeV \cite{hhhcompositionspectrum}. Validating these interpretations with LHAASO measurements requires a stable absolute energy scale, which is provided by the calibration developed here. Separately, LHAASO also reports the all-particle knee at $(3.67 \pm 0.05_{\mathrm{stat}} \pm 0.15_{\mathrm{sys}})$ PeV, with an uncertainty of only 5\% \cite{zhyprl}. These achievements set a new benchmark for ground-based cosmic-ray measurements.

Persistent discrepancies in cosmic ray spectral measurements highlight the critical need for model-independent energy calibration in indirect detection experiments. The Moon shadow effect addresses this gap: cosmic rays (CRs) blocked by the Moon produce a detectable deficit displaced westward by Earth's geomagnetic field (GMF) in a rigidity dependent manner. This displacement provides a physics-driven anchor for absolute energy calibration, only weakly dependent on hadronic-model assumptions. Unlike space-based experiments, the ground-based LHAASO-KM2A relies entirely on this method, using the shadow's displacement to validate energy reconstruction and its sharpness to assess angular resolution. It offers an unique method to minimize MC-induced biases. 
However, the strength of GMF enables us to employ this method below 300 TeV, beyond which the displacement becomes negligible compared to the pointing accuracy. This methodology was originally proposed by Clark \cite{clark1957arrival} and has since been widely adopted in experiments investigating extensive air showers (EAS) generated by high energy CRs interacting with the atmosphere. The Moon shadow was first observed by the CYGNUS collaboration in 1991 \cite{Alexandreas1991}, later by other ground-based observatories including CASA \cite{Borione1994}, ARGO-YBJ \cite{Bartolimoonshadow}, HAWC \cite{hawcmoonshadow2013}, AS$\gamma$ \cite{Tibetmoonshadow} and LHAASO \cite{aharonian2021calibration}. Additionally, continuous Moon shadow monitoring offers robust checks on the stability of these calibration metrics over time. The Moon shadow also provides valuable insights into the antiproton content of primary CRs. The opposite deflections of protons and antiprotons enable the measurement of the antiproton flux at TeV energies \cite{ppbarflux}. 

The paper is organized as follows. After the introduction, we describe the LHAASO-KM2A detector, including its configuration and performance in \autoref{Sec:detector}.
In \autoref{sec:mc}, based on a MC study, we propose the event selection criteria and the event reconstruction methodology, followed by the development of a nearly composition independent cosmic ray energy estimator for the analysis. In \autoref{sec:moonshadow}, we discuss the methodology and the  result of Moon shadow analysis. The pointing accuracy and angular resolution of the detector is covered in \autoref{sec:5}.
In \autoref{sec:energy calibration}, the results of a MC simulation of cosmic ray propagation in the Earth-Moon system are presented, along with the measurements of the Moon shadow shifts for energy calibration, including the corresponding uncertainties and the stability of the detector over time. Finally, the conclusion of the analysis is given in \autoref{sec:conclusion}.

\section{LHAASO-KM2A Detector and its Configuration}
\label{Sec:detector}

\par The Large High Altitude Air Shower Observatory (LHAASO) is a ground-based experiment for studying EAS at 4,410 meters above sea level, where the atmospheric depth is around 600 g/\text{cm$^2$} \cite{LHAASO2019}. LHAASO uses a combination of different detectors, including electromagnetic detectors (EDs), muon detectors (MDs), water Cherenkov detector array (WCDA), and wide field-of-view Cherenkov telescope array (WFCTA), to capture various aspects of particle showers. This hybrid approach allows it to measure both the energy and composition of primary particles with high accuracy, especially through data on muon content, providing cross-validation between different types of measurements.

The kilometer square array (KM2A) of LHAASO covers an area of about 1.3 \text{km$^2$} and consists of 5,216 surface EDs and 1,188 soil-overburdened MDs. Each ED has an area of 1 \text{m$^2$}, spaced 15 meters apart, while each MD covers 36 \text{m$^2$}, with a spacing of 30 meters in a triangular grid. The EDs are plastic scintillation detectors with 98\% detection efficiency for relativistic charged particles (MIPs; primarily \(e^\pm\) and also \(\mu^\pm\)) and a time resolution of 2 ns \cite{EDPerformance}. They detect electromagnetic signals from charged particles and gamma rays, providing crucial information to estimate the energy, direction, and core position of primary particles. The MDs, high-purity water Cherenkov detectors, are designed to efficiently detect muons above 1 GeV with 95\% efficiency with a time resolution of 10 ns \cite{MDproto}. Together, these detectors measure both electromagnetic particles and muons in air showers, enabling LHAASO to detect particles from tens of TeV to hundreds of PeV \cite{LHAASO2019}, while the combined ED and MD data allow KM2A to precisely reconstruct extensive air showers, enhancing the accuracy and reliability of event analysis.

 \begin{figure}[t]
    \centering
\includegraphics[width=0.9\linewidth]{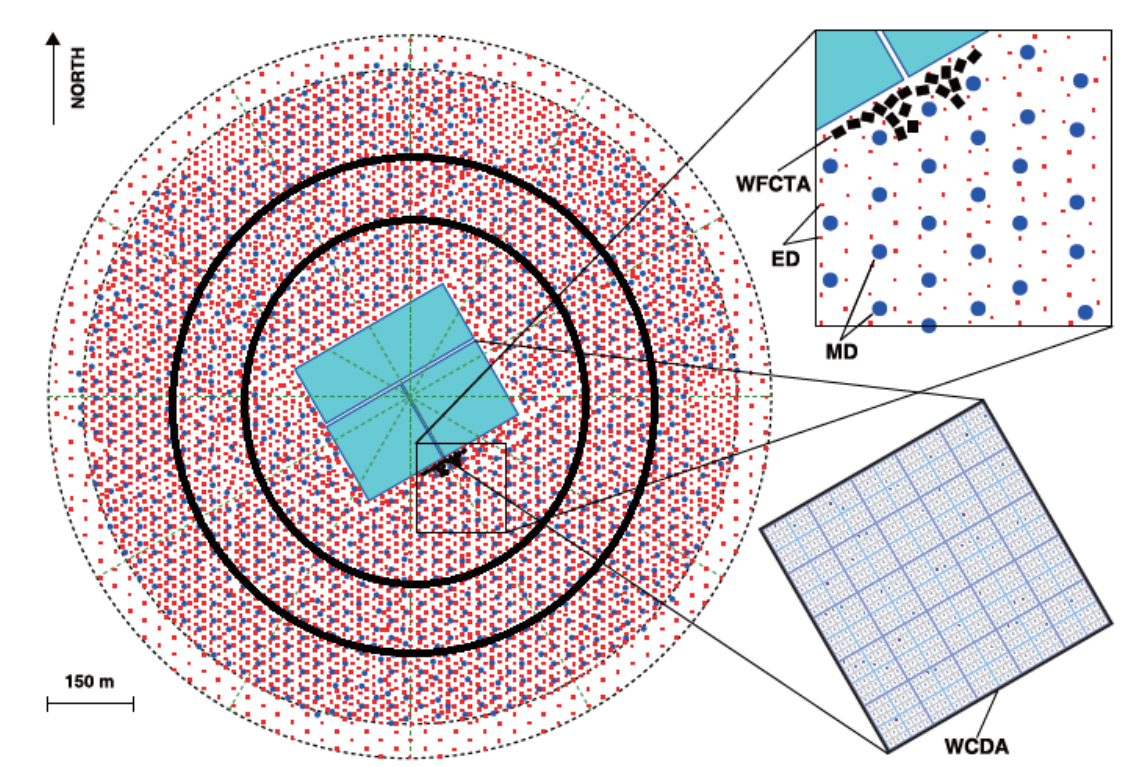}
    \caption{Layout of the LHAASO. The central square shows the WCDA (three pools). Surrounding it is the KM2A, consisting of 5,216 EDs; red dots and 1,188 underground MDs; blue circles. The WFCTA; black sectors are placed around the array and point toward the sky. The concentric black circles demonstrates only events with reconstructed shower core inside this region are retained for our analysis.}
 \label{ppr/fig:lhaaso_layout}
\end{figure}

\section{Monte Carlo Simulation}
\label{sec:mc}
A detailed Monte Carlo simulation has been performed to propagate the CRs in the Earth-Moon system.
Air showers for cosmic ray simulations were generated using CORSIKA (v77410) \cite{corsika}, incorporating the QGSJETII-04 \cite{ostapchenko2013qgsjet} and EPOS-LHC \cite{pierog2015epos} models for high energy hadronic interactions, FLUKA for low-energy interactions \cite{battistoni2015overview}. Five primary cosmic ray components (Protons, Helium, CNO, MgAlSi and Iron) were modeled over an $E^{-2}$ energy spectrum from $10^{12}$ eV to $10^{16}$ eV with an isotropic angular distribution. The simulation covered zenith angles from $0^{\circ}$ to $40^{\circ}$ and azimuth angles from $0^{\circ}$ to $360^{\circ}$ within a 1000 m radius. We simulated the detector response using G4KM2A \cite{G4KM2A}, a KM2A-specific package built on the Geant4 framework, which propagates secondary particles to ground and simulates the array response, with minimum of 10 ED triggers within 200 m after noise filtering. The total number of simulated events for CRs is $5.55 \times 10^8$. After applying the above-mentioned criteria, the total number of events for the analysis is $1.02 \times 10^7$, while cosmic-ray events were weighted based on various composition models, including Gaisser H3a \cite{gsrmodel}, Hörandol \cite{Horandel2005}, GSF \cite{gsfmodel} and GST \cite{GSTmodel} models.

\subsection{Event Selection Criteria}
\label{subsec:event selection}
The event selection criteria are identical to those outlined in \cite{zhyprl}, with exceptions regarding the number of electromagnetic particles $N_e$ and ED hits in our case. These criteria ensure good data quality and are summarized as follows:

\begin{enumerate}
\item To capture cosmic ray showers near their maximum within a range of two energy decades, zenith angles ranging from $10^\circ$ to $30^\circ$ are selected, corresponding to atmospheric depths of 610–690 $\mathrm{g/cm^2}$.
\item The location of the reconstructed shower core is restricted to inner and outer ring radii of 320 m and 420 m, respectively, as shown in \autoref{ppr/fig:lhaaso_layout} (black circles). This ensures that the core of the cosmic ray shower is far from the edges of the KM2A array, preventing loss of muons or electromagnetic particles within the range of 40-200 m. This selection minimizes the possibility of erroneously reconstructing showers with cores outside the array.
\item The number of muons $N_\mu$ must be greater than 0, while the number of electromagnetic particles $N_e$ must be greater than 20, with at least 20 EDs fired. Both $N_\mu$ and $N_e$ are selected within the range of 40–200 m from the shower core.
\end{enumerate}

\begin{figure}[ht]
    \centering
    \includegraphics[width=0.9\linewidth]{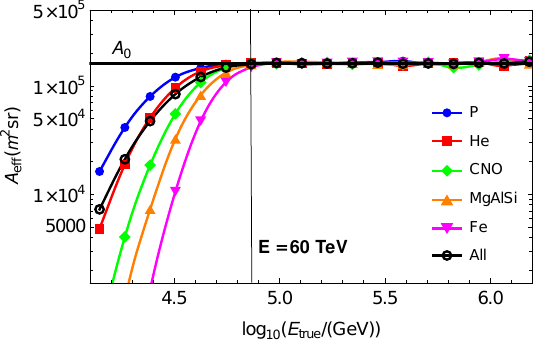}
    \caption{Bin-wise aperture ($A_{\textrm{eff}}$) of the KM2A full-array for cosmic ray showers vs. the primary energy of each mass group, calculated with the EPOS-LHC interaction model. ``All'' represents the weighted values of all components according to the Gaisser model. The black horizontal line shows the calculated aperture $A_{0} = 0.16\, \text{km}^2\text{sr}$. The vertical black line marks the energy above which the detector is fully efficient for ``All''. 
}
    \label{ppr/fig:km2a_efficiency}
\end{figure}

\begin{figure*}[htb]
\centering
\includegraphics[width=0.9\linewidth]{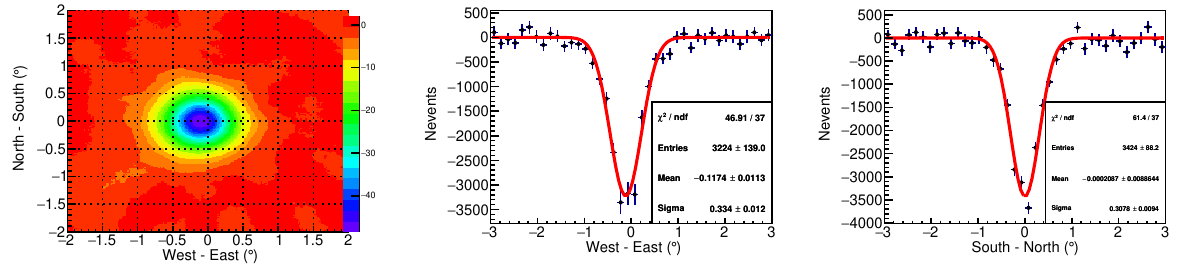} 
\caption{Left: Significance map of the Moon region for events satisfying the event selection criteria in \autoref{subsec:event selection}, observed by the LHAASO-KM2A for the period July 21-Jan 25 (1290) days. Middle: The shift along the Right-Ascension (RA) caused by the Moon shadow. Right: The shift along the Declination (DEC) direction.}
\label{ppr/fig:1DMoon_shadow}
\end{figure*}

With the above data selection, the detector performance is monitored over a wider energy range, but our focus is on the 20 TeV to 300 TeV interval, where the energy calibration is performed. As shown in the \autoref{ppr/fig:km2a_efficiency}, the detector achieves almost full efficiency for energies above 60 TeV. 

\subsection{Energy Estimator}
\label{subsec:estimator}
Given that the cosmic ray flux for all particle spectra depends on the composition, it is essential to examine the response of different energy estimators before selecting one for energy calibration. Traditional energy estimators like $N_{\text{size}}$ based on the modified NKG function and $\rho_{50}$, which measures particle density at 50 m from the shower core, show a clear dependency on cosmic ray composition \cite{zhang2022approaches}. Thus, they lack in accuracy of energy reconstruction, primarily due to the reason that they exclude muonic content of the shower, which is a key to explore cosmic ray composition. The most promising estimator emerges from the combination of electromagnetic and muonic content, defined by $N_{e\mu} = N_e + 2.8N_\mu$. 

The simulation results confirm that this energy estimator is effectively composition-independent at high energies $E \geq 300$ TeV \cite{zhang2022approaches}, making it preferable for reducing bias in
energy reconstruction, but we cannot calibrate the detector at such higher energies. In this work, we evaluate its response over 20–300~TeV using six bins mentioned in \autoref{tab:moon_shadow}. For the first three bins (median energies $\lesssim 50$~TeV), the response of this energy proxy is non-linear, primarily because the detector is not yet fully efficient and the estimator retains residual composition-dependence. At higher energies ($\gtrsim 60$~TeV), as the detector reaches full efficiency shown in \autoref{ppr/fig:km2a_efficiency}, the response becomes approximately composition-independent.

\section{Moon Shadow Analysis}
\label{sec:moonshadow}
To analyze the Moon shadow, three sky maps in celestial coordinates (right ascension (R.A) and declination (DEC)) are constructed: an event map, which is centered on the Moon, spans $10^{\circ}\times10^{\circ}$, a background map both with a bin size of $0.02^{\circ}\times0.02^{\circ}$ and a significance map to determine statistical relevance. Accurate background estimation is essential to separate the deficit signal from fluctuations. Several methods are widely used: the Equi-zenith angle method \cite{equizenith}, the Surrounding windows method \cite{surroundingwindow}, the Direct integral method \cite{abdo2012observation}, and the Time-Swapping method \cite{Fleysher2004}. In this analysis, we favor the equi-zenith angle method for its robustness. This approach involves averaging cosmic ray events from off-source windows that match the on-source window in the same size, zenith angle, and time intervals, effectively reducing distortions from environmental and instrumental fluctuations such as changes in pressure and temperature. The background estimation is conducted in the reference frame of the experiment using local coordinates (zenith angle ($\theta$) and azimuth ($\phi$)). We used a total of 11 windows, among them ten (10) off-source windows symmetrically aligned on both sides of the on-source window at the same zenith angle. When $\theta \geq 18^\circ$, the window size remains fixed at $5{^\circ} \times 5{^\circ}$. However, when the zenith angle of the Moon falls below this threshold ($\theta < 18^\circ$), the windows begin to overlap. To address this, we introduce a variable window radius that dynamically adjusts to values smaller than $5^{\circ}$, preventing event duplication across adjacent windows and ensuring each event is captured uniquely. 
 
The significance map tells us the statistical significance of detected events. This map is obtained by comparing the number of events from the event map and the background map. In accordance with the Gaussian PSF (Point Spread Function) properties, smoothing is applied to the event and background maps before evaluating the per-bin significance to reduce statistical fluctuations by averaging neighboring bins, improving the signal-to-background ratio and creating a clearer, more stable map. The angular resolution is then used to compute the overall significance, providing a refined measure of statistical confidence for detected events under the Gaussian weighting. The deficit significance in each source-map bin, relative to the corresponding background bin, is computed with the Li–Ma statistic (\autoref{eq:Li_Ma}), following Eq.~(17) of Ref.~\cite{liandma1983}.


\begin{equation}
\hspace*{-\leftmargin}  
\resizebox{\columnwidth}{!}{$
S = \sqrt{2} \left\{ N_{\text{on}} \log{\left[\frac{1+\alpha}{\alpha} \left(\frac{N_{\text{on}}}{N_{\text{on}} + N_{\text{off}}}\right)\right]} 
+ N_{\text{off}} \log{\left[(1+\alpha) \left(\frac{N_{\text{off}}}{N_{\text{on}} + N_{\text{off}}}\right)\right]} \right\}^{\frac{1}{2}}
$}
\label{eq:Li_Ma}
\end{equation}
where $N_{\text{on}}$ and $N_{\text{off}}$ are the number of events in the off-source and on-source regions, respectively, and $\alpha$ is the ratio between off-source and on-source regions.

Finally, the finite angular size of the Moon contributes to the observed spread of the signal; therefore, we must account for this effect when measuring the detector's angular resolution using \autoref{eq:angular_resolution}.


\begin{equation}
\text{RMS} = \sigma_{\theta} \sqrt{1 + \left( \frac{r_m}{2\sigma_{\theta}} \right)^2}
\label{eq:angular_resolution}
\end{equation}
where $\sigma_{\theta}$ is the angular resolution of the detector, $r_m$ is the angular radius of the disc of the Moon ($\approx 0.252^\circ$), and $\text{RMS}$ is the width of the observed signal, considering both the detector's PSF and the Moon's finite size.

\autoref{ppr/fig:1DMoon_shadow} shows, the significance map obtained for the Moon shadow observed with data recorded from 21 July 2021 to 31 Jan 2025 (1290 days) and about 2029.6 hours on-source time with the $1.247 \times10^8$ Moon shadow events, which satisfied our event selection criteria as mentioned in \autoref{subsec:event selection}. The maximum statistical significance obtained for this total analysis is 48$\sigma$.

\section{Pointing Accuracy and Angular Resolution}
\label{sec:5}

\begin{figure}[H]
    \centering
      \includegraphics[width=0.36\textwidth]{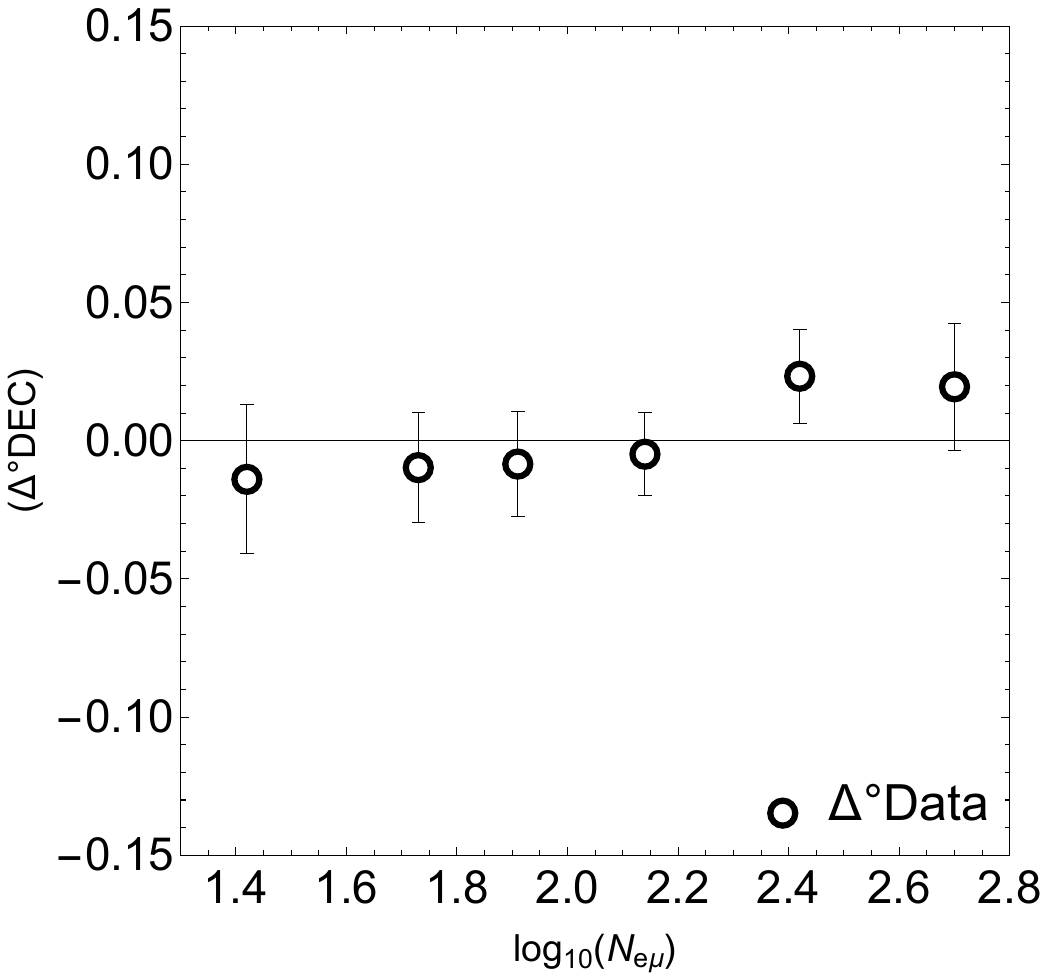}
    \caption{Bin-wise pointing accuracy of the detector.}
    \label{ppr/fig:radecshift}
\end{figure}

The pointing accuracy of the detector is evaluated by observing the Moon's shadow, with the accuracy derived from the shifts in the North-South direction (DEC), which are minimally affected by the Galactic Magnetic Field (GMF). As shown in \autoref{ppr/fig:1DMoon_shadow}, the observed Moon shadow in the North--South direction shows no measurable displacement, confirming the detector's exceptional pointing accuracy. The derived pointing accuracy is $0.0002^\circ \pm 0.009^\circ$, with the dominant uncertainty being $0.009^\circ$, so we used this factor for reporting pointing error, which is significantly smaller than the $0.1^\circ$ benchmark reported by the LHAASO-KM2A collaboration.
Bin-wise data validation further endorse these results in \autoref{ppr/fig:radecshift}.

The angular resolution of the detector was obtained bin-wise, as shown in \autoref{tab:angular} and it was further assessed via total analysis along the DEC direction as shown in \autoref{ppr/fig:1DMoon_shadow}. After correction for the Moon's disk geometry, the derived PSF function is \(\sigma_{\mathrm{PSF}} = 0.275^\circ \pm 0.009^\circ\). These measurements are consistent with the $0.29^\circ$ benchmark reported by the Ref.~\cite{cao2025data}.

\setlength{\tabcolsep}{8pt}
\begin{table}[H]
\centering
\begin{tabular}{c c c}
\hline
\textbf{$\log_{10}(N_{e\mu})$} &
\textbf{$\sigma^\circ_{RA}$} & \textbf{$\sigma^\circ_{DEC}$} \\
\hline
$1.2-1.64$ & 0.42 $\pm$ 0.028 & 0.41 $\pm$ 0.028 \\
$1.64-1.82$ & 0.34 $\pm$ 0.021 & 0.34 $\pm$ 0.028 \\
$1.82-2.0$ & 0.33 $\pm$ 0.02  & 0.3  $\pm$ 0.021 \\
$2.0-2.28$ & 0.26 $\pm$ 0.015 & 0.25 $\pm$ 0.017 \\
$2.28-2.56$ & 0.24 $\pm$ 0.017 & 0.2  $\pm$ 0.017 \\
$2.56-2.84$ & 0.2  $\pm$ 0.018 & 0.18 $\pm$ 0.022 \\
\hline
\end{tabular}
\caption{Bin-wise $\sigma^\circ_{RA}$ and $\sigma^\circ_{DEC}$ without Moon's disc correction. The max significance and shift values around RA \& DEC are mentioned in \autoref{tab:moon_shadow}.}
\label{tab:angular}
\end{table}

\begin{figure*}[ht]
    \includegraphics[width=0.3\textwidth]{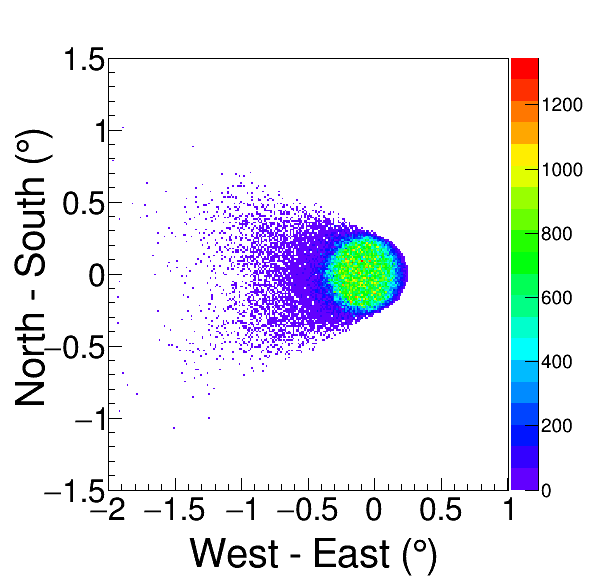}
     \includegraphics[width=0.3\textwidth]{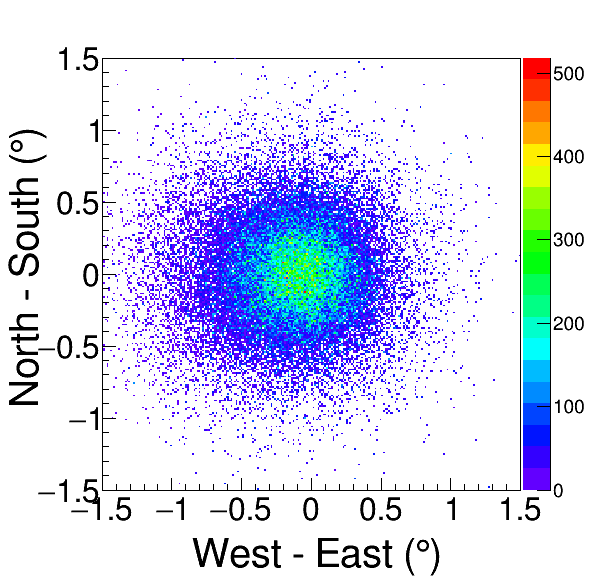}
    \includegraphics[width=0.3\textwidth]{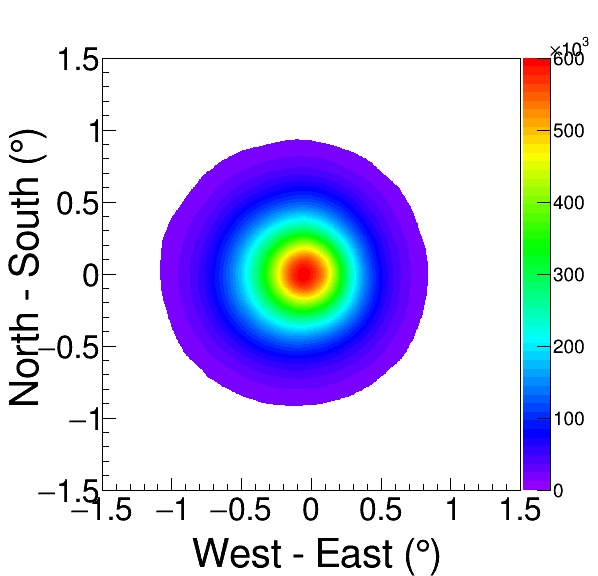}
    \caption{The effect of folding different contributions to the Moon signal. Left: effect of the GMF on an ideal detector without PSF. Middle: effect of the detector PSF with the GMF. Right: effect of smoothing procedure after applying GMF. The color scale represents the number of showers lying on the single pixel of the figure.}
    \label{ppr/fig:ray_tracing}
\end{figure*}

\section{Energy Scale Calibration of LHAASO-KM2A}

\label{sec:energy calibration}
The observed shift in Moon shadow for each $N_{e\mu}$ bin represents an average effect caused by the deflection of all cosmic ray particles in the Moon’s direction and belonging to the given bin. The dependence of the deflection on both the energy and charge of the primary particle implies that the relationship between the measured bin-wise shift and the corresponding energy involves a composition-dependent factor. To estimate the value of this factor for each $N_{e\mu}$ bin, we must simulate the propagation of cosmic ray particles using a technique known as ray-tracing.

\subsection{Simulation of Moon shadow shifts by ray tracing (RT) using GMF}

To accurately simulate the Moon shadow effect, it is crucial to consider the magnetic environment between the Moon and Earth. The region is primarily influenced by the geomagnetic field (GMF) and the interplanetary magnetic field (IMF) carried by the solar wind, as the Moon’s intrinsic magnetic field is effectively zero. However, the GMF dominates, particularly in the near-Earth region, where most of the deflection occurs \cite{zhamin}. So, to estimate total deflection, we use the standard empirical Moon–shadow deflection relation adopted by LHAASO \cite{aharonian2021calibration}, which is valid for events with zenith angle
$\theta \lesssim 45^\circ$ and energies above a few~TeV. Under this assumption, the deflection angle can be expressed as

\begin{equation}
    \Delta = -1.59^\circ \frac{Z}{\text{E(TeV)}}
\label{eq:<1.59Z>}
\end{equation}

However, for a more accurate simulation of particle propagation in the Earth-Moon system, the Tsyganenko-IGRF (T-IGRF) model~\cite{Tigrf} is adopted. This model incorporates both internal geomagnetic contributions and external magnetospheric effects using empirical data derived from direct observations. 
To estimate the shift for events in each $N_{e\mu}$ bin used in the Moon shadow analysis, we used the corresponding data of events from KM2A-MC that satisfied the event selection criteria outlined in \autoref{subsec:event selection} and applied the ray-tracing method \cite{Bartolimoonshadow} using the true energy of the primary cosmic ray particles. Ray-tracing is applied over one complete Moon cycle, with a 10 second time bin to track its changing position.  Events were generated within a circular window of radius $5^\circ$ around the Moon using the T-IGRF model and only those blocked by the Moon’s disk were retained for analysis. A total of approximately 4 million events were simulated, of which only 52 thousand events were blocked by Moon. A 2D map of the deficit count of events for all values of $N_{e\mu}$, shown in the left plot of \autoref{ppr/fig:ray_tracing}, clearly exhibits a westward shift due to GMF.

\begin{figure}[H]
    \centering
\includegraphics[width=0.32\textwidth]{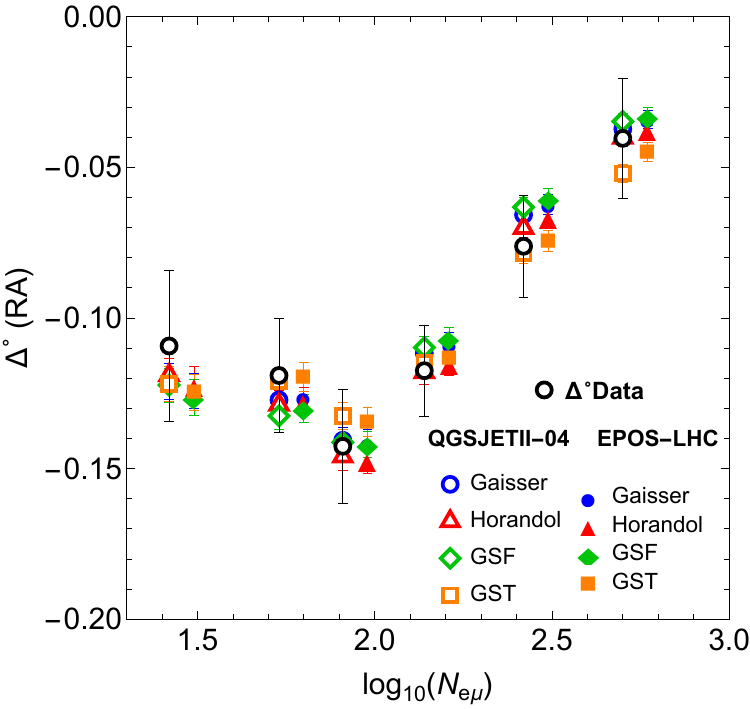}
    \caption{RA shifts per $N_{e\mu}$ bin from ray-tracing simulations (two interaction and four composition models), with EPOS-LHC points slightly offset for clarity and data-derived shifts overlaid for comparison.}
\label{ppr/fig:RAshift}
\end{figure}

To incorporate the effect of finite angular resolution of the detector, we convolve the 2D deficit count map with a bin-specific Gaussian point-spread function (PSF). We use six $\log(N_{e\mu})$ bins; each bin corresponds to a different median energy and has its own Gaussian PSF whose width $\sigma$ is set by the angular resolution in that bin. The resultant map is further subjected to the same smoothing procedure, as applied on the significance map in the Moon shadow analysis, to obtain a clear picture of shifts along RA and DEC. However, RA/DEC shifts are always extracted from the PSF–applied maps. The shifts are obtained by constructing 1D projections of the PSF–applied deficit maps and fitting a 1D Gaussian to the deficit profile. In the middle and right panels of \autoref{ppr/fig:ray_tracing}, we show the effect of applying the detector's PSF and the smoothing. In simulations, the RA shifts for each $N_{e\mu}$ bin are obtained as described and shown in \autoref{ppr/fig:RAshift}. Considering the East–West deflection, we observe good agreement between data and simulations for the four composition models (Gaisser, H\"orandol, GSF, and GST) and two interaction models (QGSJETII-04 and EPOS-LHC), respectively.

\begin{figure*}[ht]
  \centering
  \includegraphics[width=0.4\textwidth]{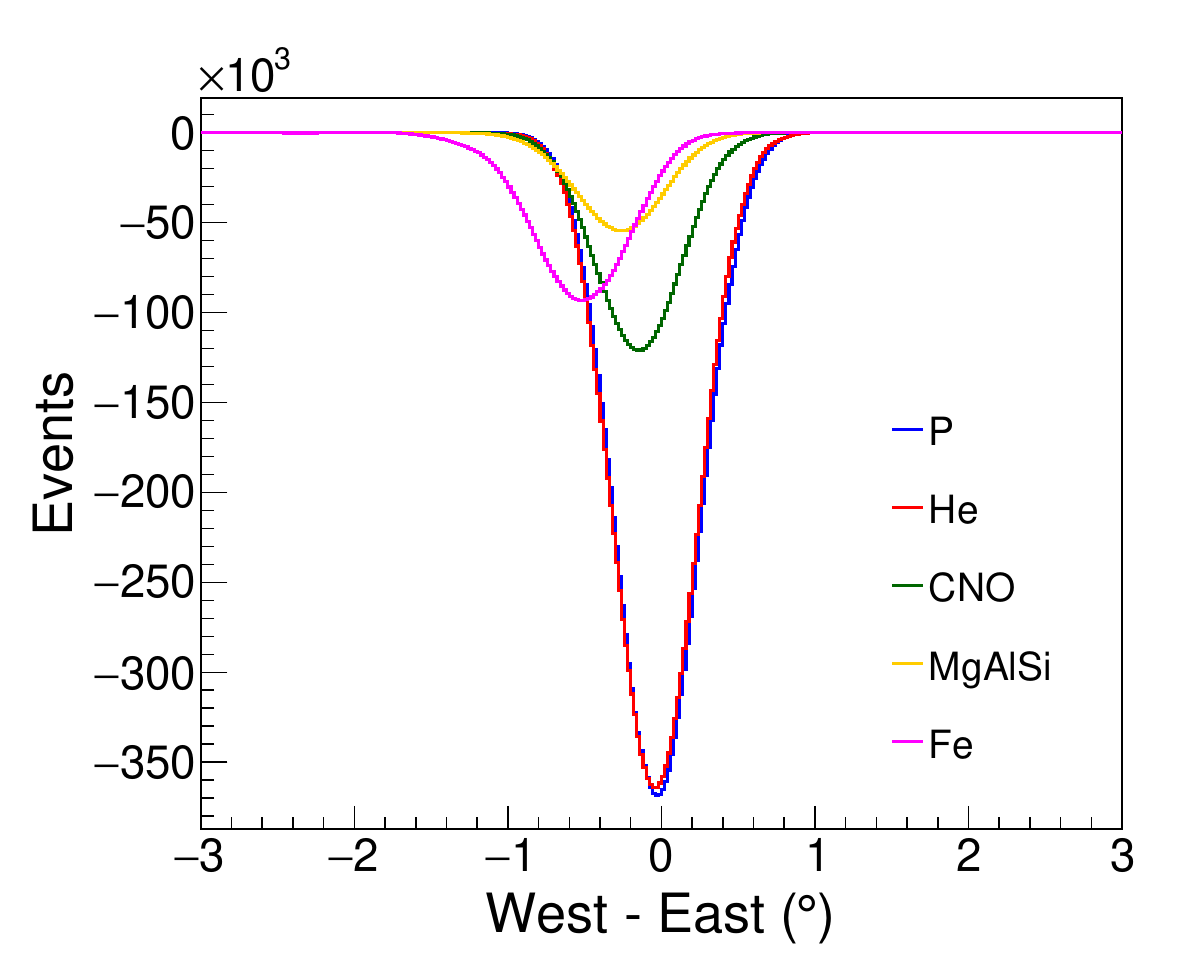}  \includegraphics[width=0.4\textwidth]{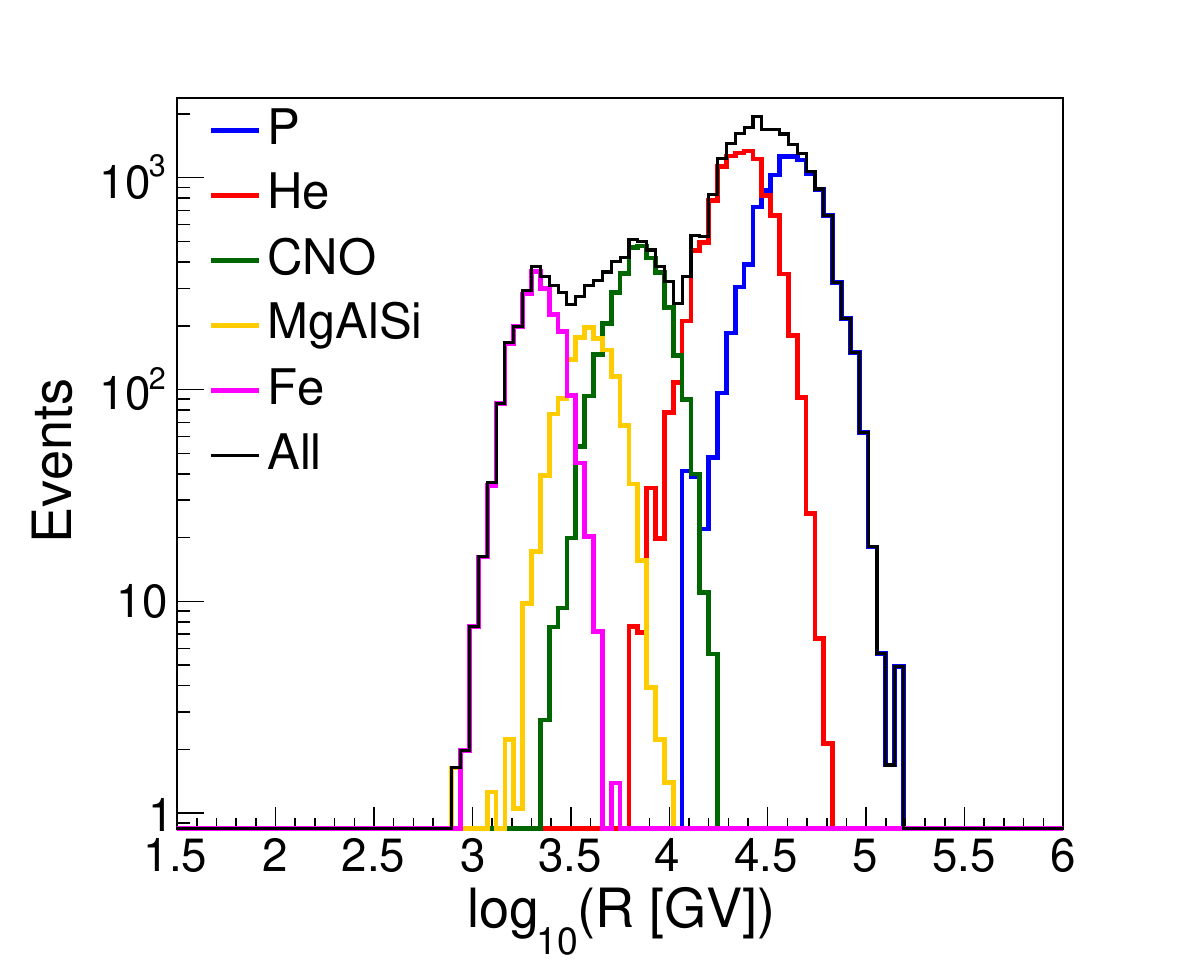}  
  \caption{QGSJETII-04 with Gaisser model is used for this graph. Left: the contribution of different primaries to the Moon shadow deficit (excluding All for better visualization of results for higher Z).
    Right: rigidity distribution for events induced by different nuclei.}
  \label{ppr/fig:deficit_rt}
\end{figure*}

A value of simulated shift along RA is an outcome of the contribution of different cosmic-ray primaries including P, He, CNO, MgAlSi, and Fe. In the left panel of \autoref{ppr/fig:deficit_rt}, we show these contributions separately in the 1D projection map of deficit counts of events belonging to the 3rd $N_{e\mu}$ bin. The results are consistent with \autoref{eq:<1.59Z>}, requiring that the shift is proportional to $Z$ of the primary. The same is also highlighted in the right panel of \autoref{ppr/fig:deficit_rt}, showing the rigidity ($R=\mathrm{E(TeV)}/Z$) distribution of events induced by different primaries.

The value of the simulated shift $\Delta_{i(\mathrm{RT})}$ along RA, together with the corresponding median energy $E_{i(\mathrm{med})}$ for each $N_{e\mu}$ bin, is used to determine the effective value of $\langle Z\rangle_i$ according to the following formula:
\begin{equation}
\Delta_{i(\mathrm{RT})} = -1.59^\circ\,\frac{\langle Z\rangle_i}{E_{i(\mathrm{med})}}
\label{ithmedianZ}
\end{equation}
Although the East–West deflection decreases with increasing energy at fixed charge, the first two bins in \autoref{ppr/fig:RAshift} show a smaller average deflection because KM2A is not fully efficient below 60~TeV (see \autoref{ppr/fig:km2a_efficiency}) and the $N_{e\mu}$ response is also non-linear, which biases the accepted events toward lighter primaries and thus lowers the effective $\langle Z\rangle$.

\begin{figure}[t]
    \centering
\includegraphics[width=0.4\textwidth]{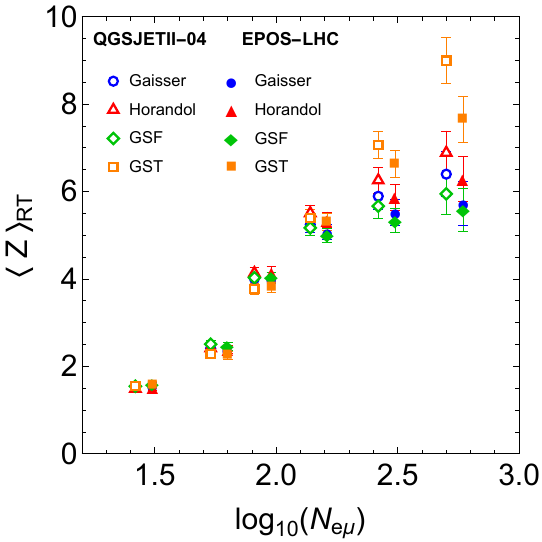}
    \caption{Bin-wise values $\langle{Z}\rangle_i$ from the RT (ray-tracing) method using two interaction and four composition models. EPOS-LHC values are slightly offset for clarity. Error bars propagated random errors based on ray-tracing shifts.}
\label{RT_mz}
\end{figure}

We use the \autoref{ithmedianZ} to determine the bin-wise values of $\langle{Z}\rangle_i$ for four composition models (Gaisser, Horandol, GSF, and GST) and two interaction models (QGSJETII-04 and EPOS-LHC), respectively, and shown in \autoref{RT_mz}. The plot shows that at low energy (below 60 TeV) $\langle{Z}\rangle_i$ is virtually insensitive to composition and interaction models. Finally, we use the data of the calculated values of $\langle{Z}\rangle_i$ to estimate the systematic and statistical errors in its mean value for each $N_{e\mu}$ bin. These values are reported in the fifth column of the \autoref{tab:moon_shadow}. 

\subsection{Measurement of Moon shadow shifts with LHAASO-KM2A}

\begin{figure*}[htb]
\centering
    \includegraphics[width=0.3\textwidth]{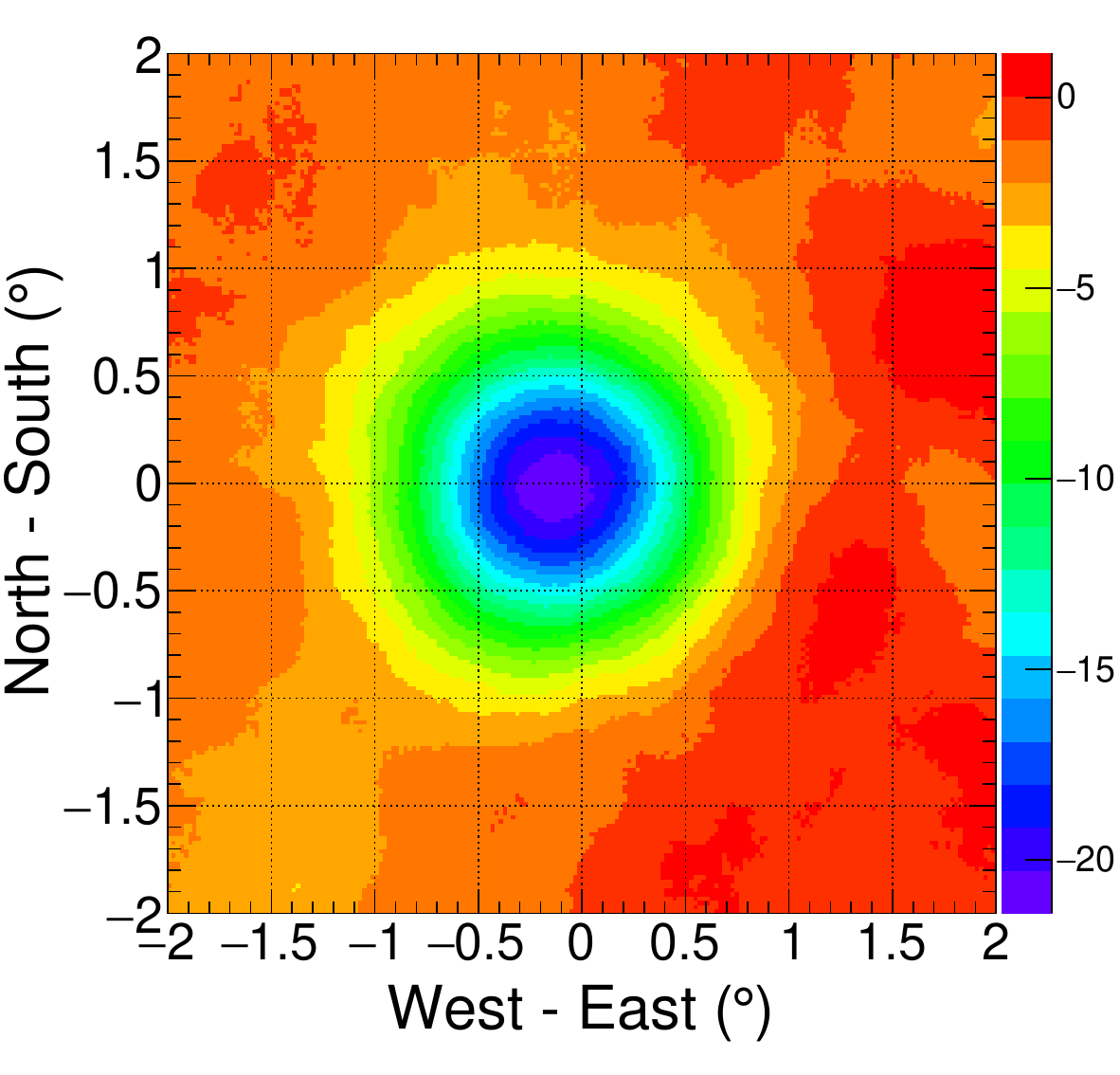}
     \includegraphics[width=0.3\textwidth]{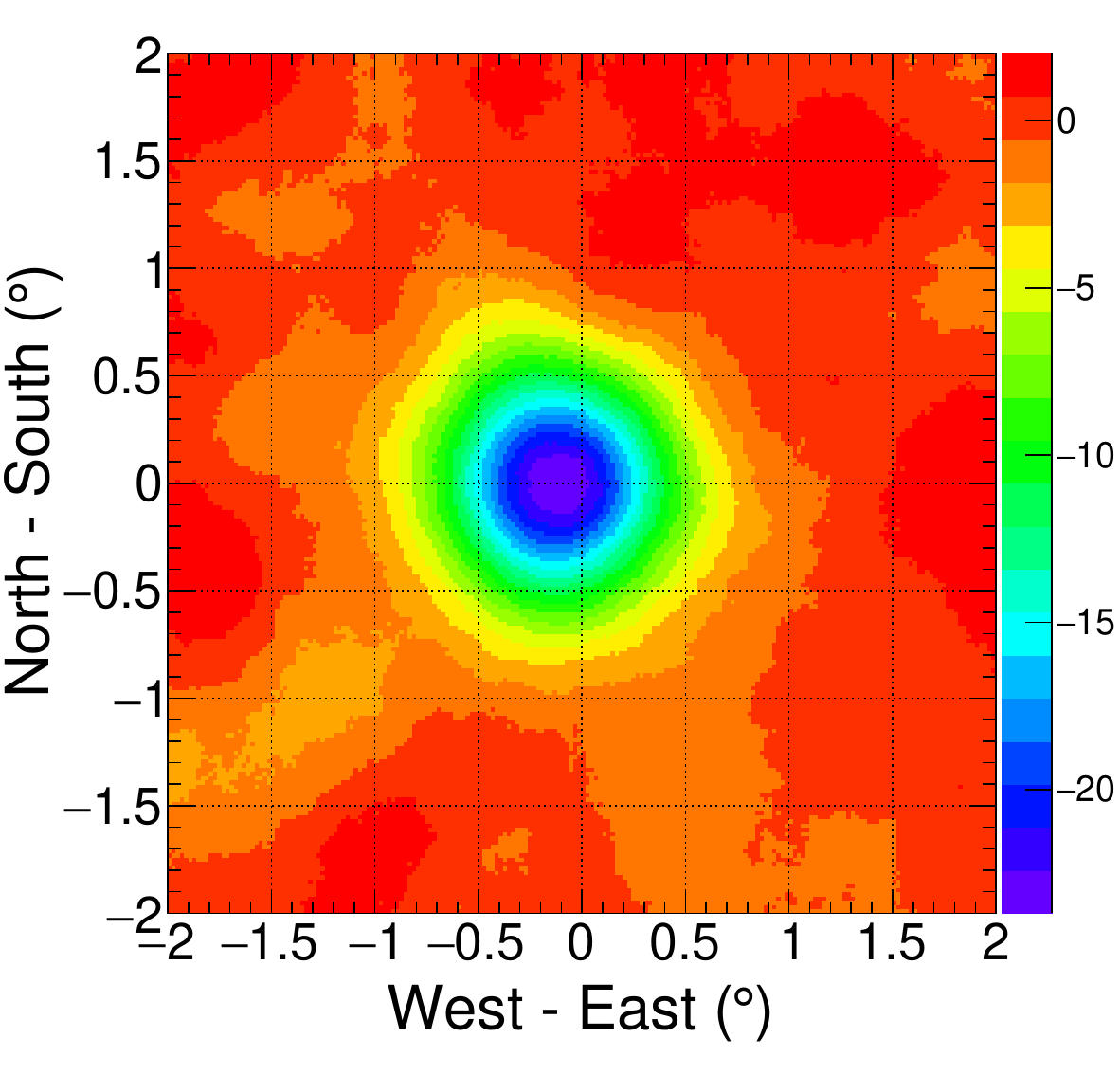}
    \includegraphics[width=0.3\textwidth]{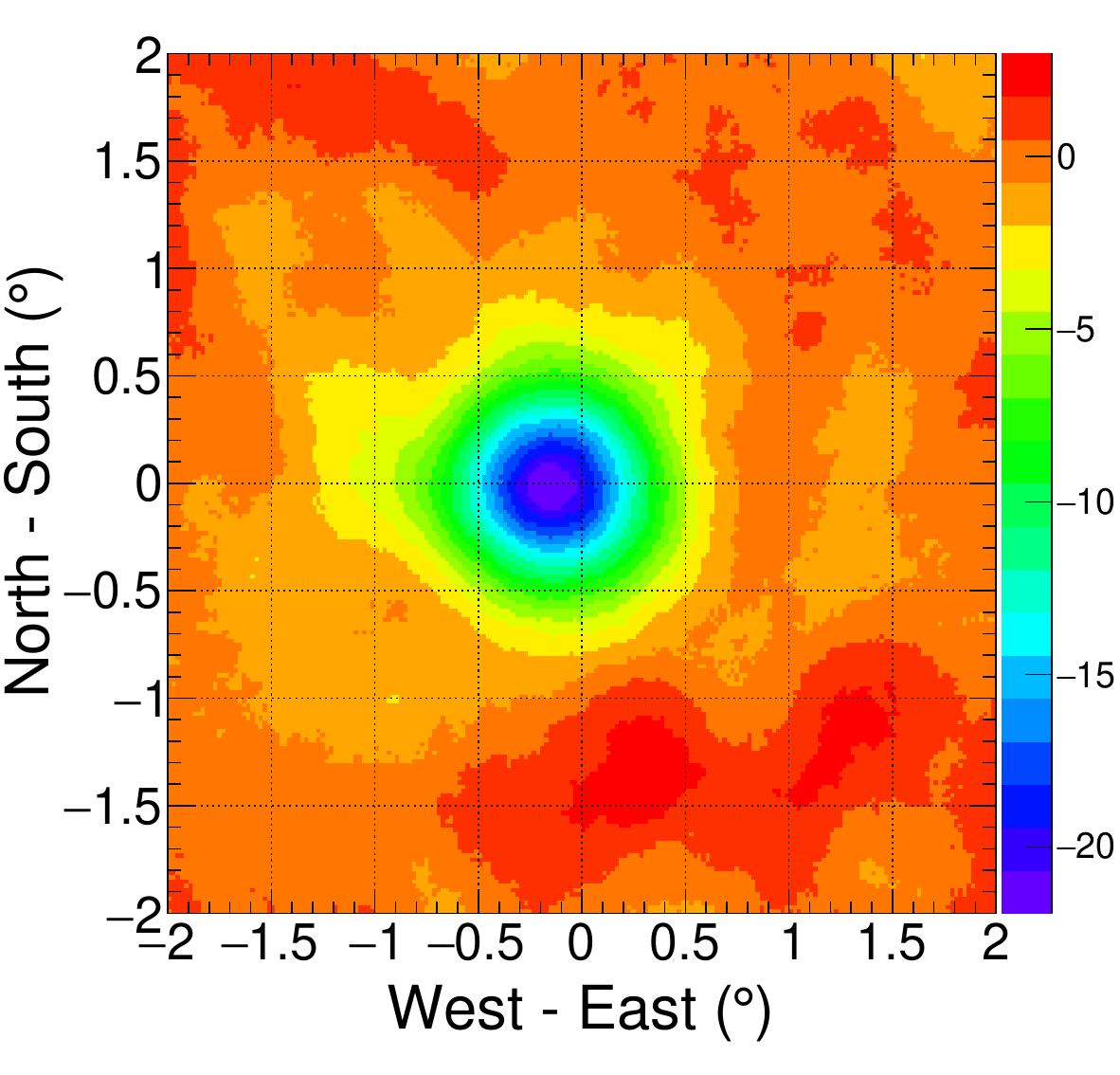}
     \includegraphics[width=0.3\textwidth]{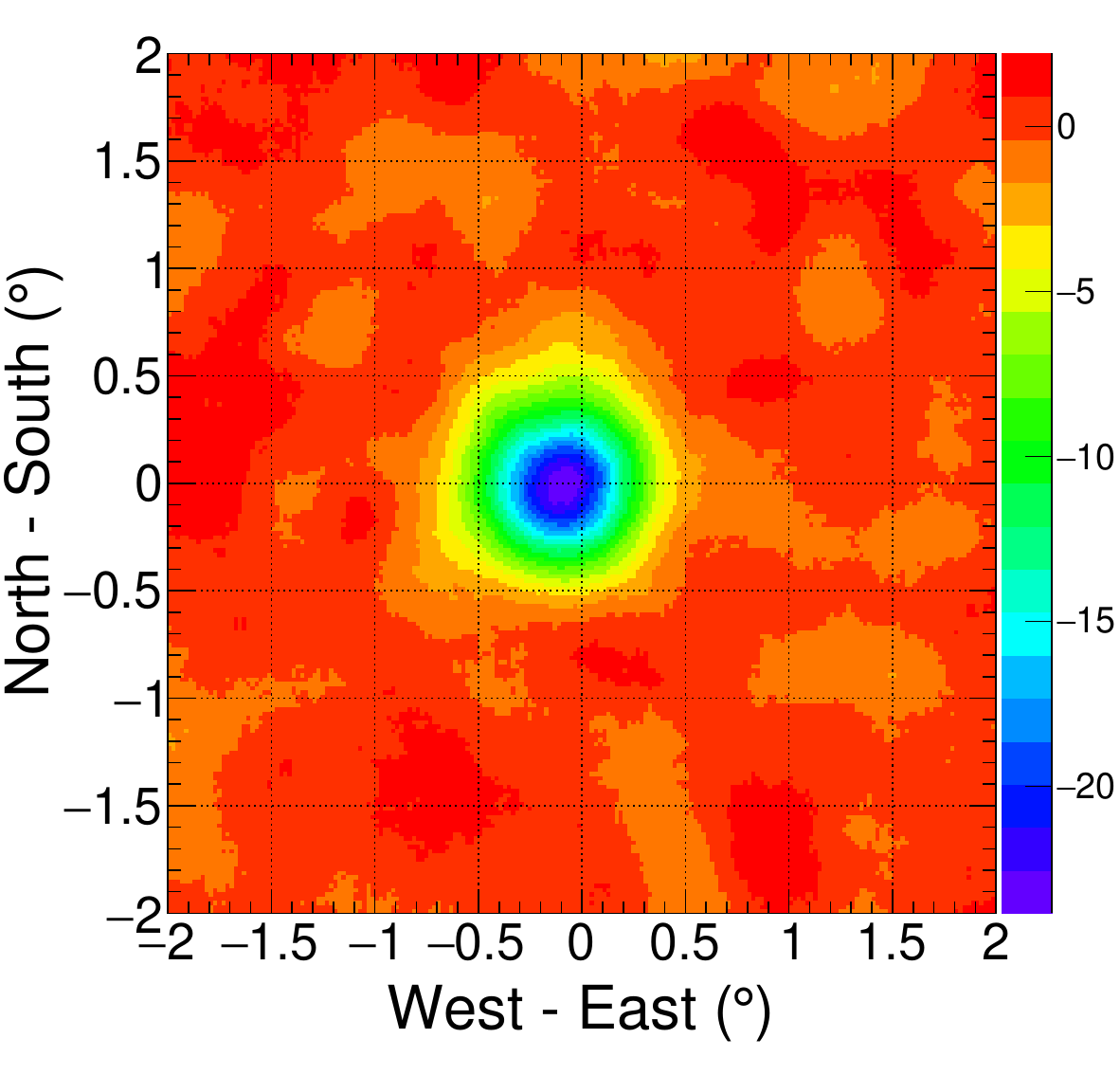}
     \includegraphics[width=0.3\textwidth]{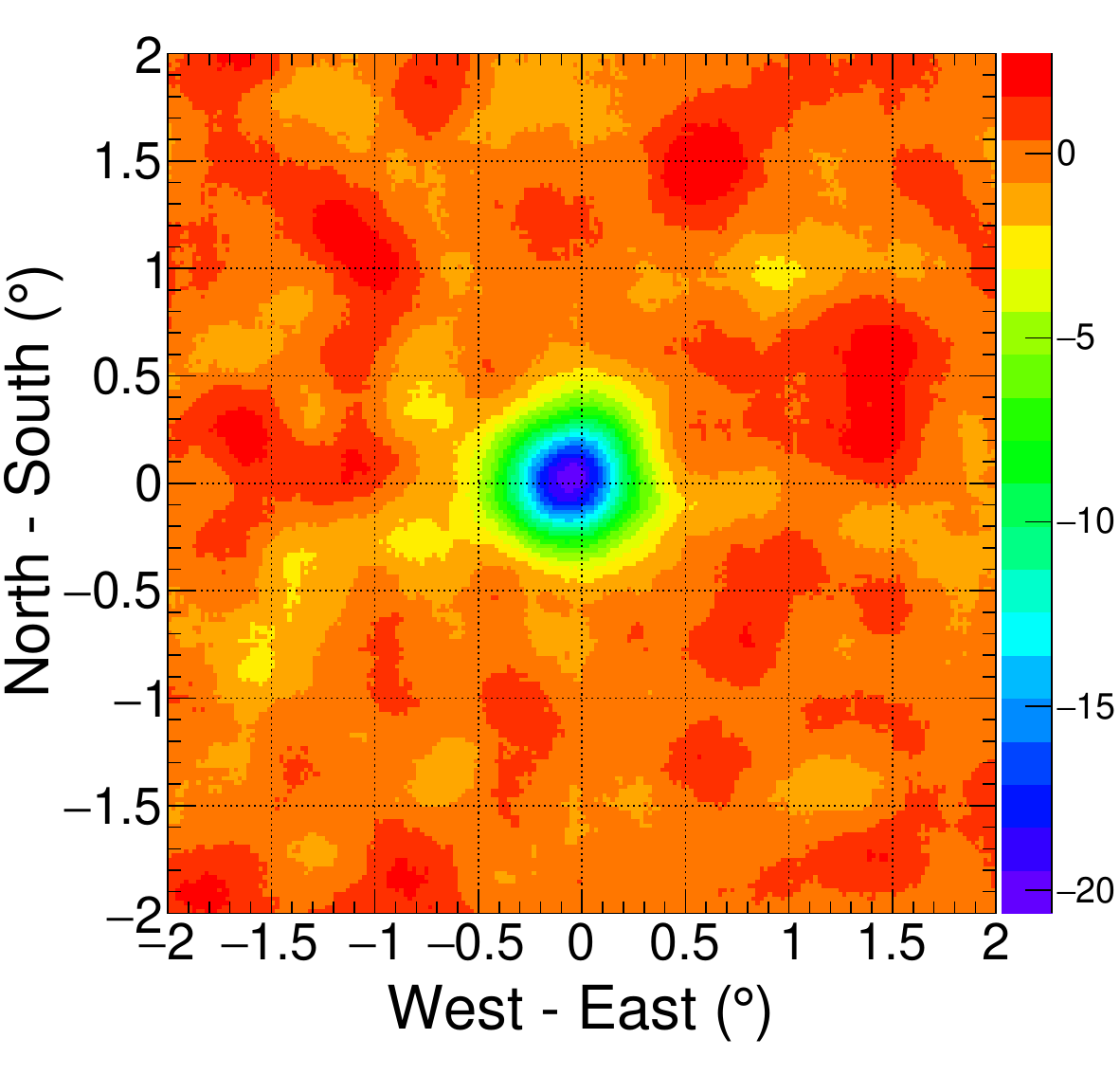}
    \includegraphics[width=0.3\textwidth]{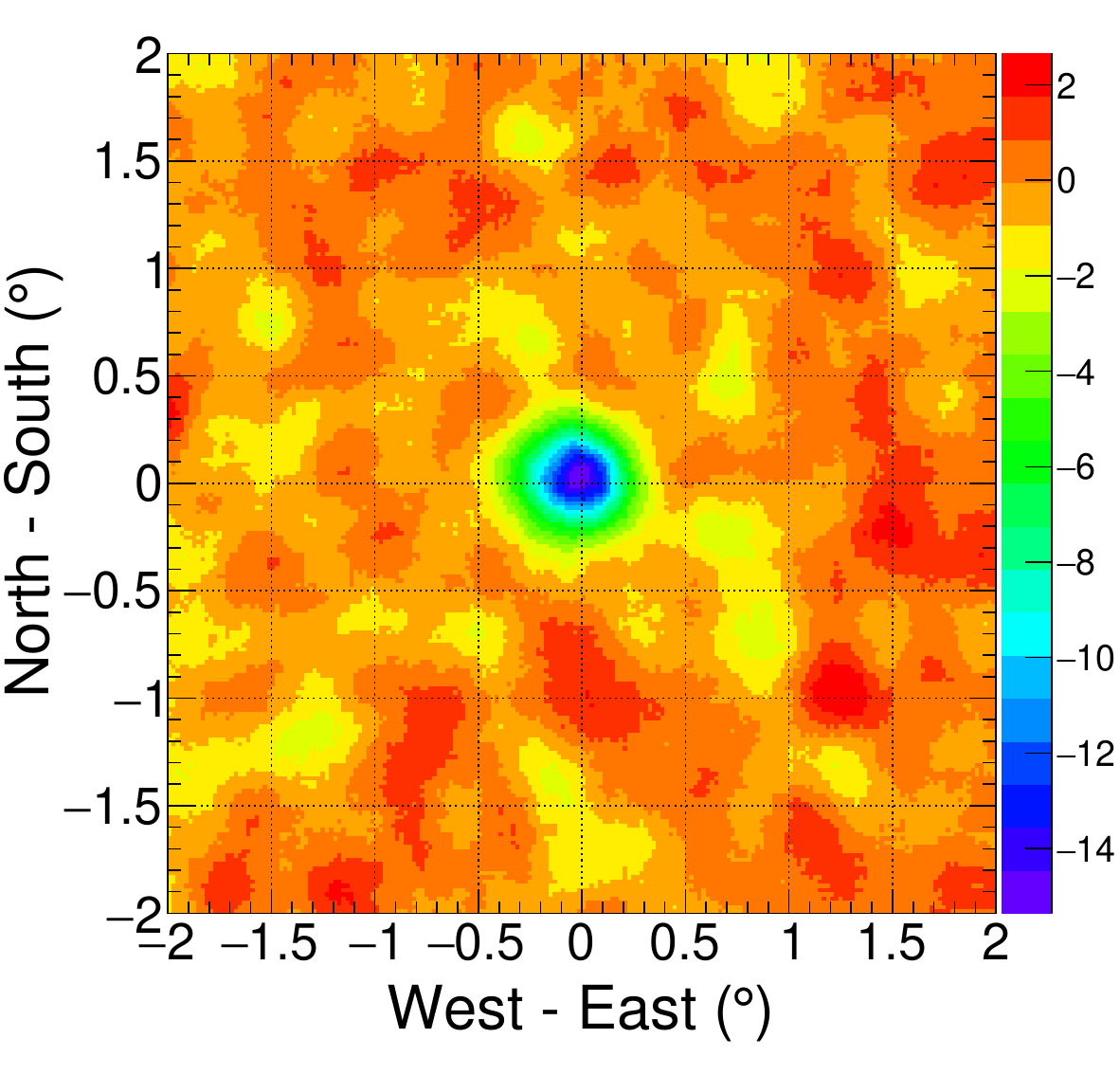}
    
    \caption{Left to Right: Bin wise Moon shadow significance maps for events satisfying the selection criteria discussed in \autoref{subsec:event selection} and the coordinates are centered on the Moon position. The color scale represents the statistical significance of the deficit in terms of the standard deviation. Bin wise significance and the median energy corresponding to each map ($6^{th}$ column) are reported in \autoref{tab:moon_shadow}.}
    \label{ppr/fig:6moon_shadow}
\end{figure*}

To analyze the deflection angle as a function of energy, cosmic ray events were divided into sub groups of  \(\log_{10}(N_{e\mu})\). For events in each bin, we construct 2D deficit and significance maps centered at Moon's position, as described in the \autoref{sec:moonshadow}. The significance maps, after applying smoothing procedure, are shown in \autoref{ppr/fig:6moon_shadow}, whereas, corresponding maximum values of significance in each map are reported in the second column of \autoref{tab:moon_shadow}.  To extract west-east and north-south shifts, we obtain 1D projections of the deficit maps, along RA and DEC direction respectively, and fit each with 1D Gaussian function $Ae^{-\frac{(x - \mu)^2}{2\sigma^2}}$, where $\mu$ and $\sigma$ correspond to the displacement and angular resolution of the detector. In the third and fourth columns of the \autoref{tab:moon_shadow},  we report the values of displacements with statistical error along RA as shown in \autoref{ppr/fig:RAshift} and DEC direction in \autoref{ppr/fig:radecshift}, respectively. We find consistent values of angular resolutions along RA and DEC as mentioned in \autoref{tab:angular}. 

\begin{table*}[htb]
\centering
\begingroup
\renewcommand{\arraystretch}{1.7}%
\resizebox{\textwidth}{!}{%
\begin{tabular}{c c c c c c c}
\hline
\textbf{Range of} & \textbf{Max.} & \textbf{Shift of the} &  
\textbf{Shift of the} &
\textbf{Model} & \textbf{RT} & \textbf{Calibrated} \\
\textbf{Estimator} & \textbf{Significance} & \textbf{Moon Shadow} & \textbf{Moon Shadow} & \textbf{Averaged value of} & 
\textbf{Median} &
\textbf{Median} \\
\textbf{$\log_{10}(N_{e\mu})$}& \textbf{($\sigma$)} & \textbf{\( \Delta^\circ_{\text{data}} \) (RA)} & \textbf{\( \Delta^\circ_{\text{data}} \) (DEC)} & 
\textbf{$\langle{Z}\rangle$} & 
\textbf{$\langle \tilde{E}\rangle$ (TeV)} &
\textbf{$E_{\text{cal}}$ (TeV)} \\
\hline
$1.2-1.64$ & -21.4 & $-0.1093 \pm 0.025$ & $-0.01399 \pm 0.027$ & $1.54 \pm (0.02)_{\text{stat}} \pm (0.03)_{\text{sys}}$ & 20 & $23_{-4}^{+7}\pm (0.6)_{\langle{Z}\rangle}$ \\
\hline
$1.64-1.82$ & -23.7 & $-0.1191 \pm 0.019$ & $-0.00975 \pm 0.02$ & $2.39 \pm (0.03)_{\text{stat}} \pm (0.08)_{\text{sys}}$  & 30 & $32_{-4}^{+6} \pm (1.2)_{\langle{Z}\rangle}$ \\
\hline
$1.82-2.0$ & -21.9 & $-0.1426 \pm 0.019$ & $-0.00847 \pm 0.019$ & $3.99 \pm (0.04)_{\text{stat}} \pm (0.13)_{\text{sys}}$ & 45 & $45_{-5}^{+7}\pm (1.5)_{\langle{Z}\rangle}$\\
\hline
$2.0-2.28$ & -23.84 & $-0.1175 \pm 0.015$ & $-0.0049 \pm 0.015$ & $5.26 \pm (0.06)_{\text{stat}} \pm (0.16)_{\text{sys}}$ & 75 & $71_{-8}^{+10}\pm (2.4)_{\langle{Z}\rangle}$ \\
\hline
$2.28-2.56$ & -20.6 & $-0.0762 \pm 0.017$ & $0.0233 \pm 0.017$ & $6.02 \pm (0.1)_{\text{stat}} \pm (0.58)_{\text{sys}}$ & 140 & $126_{-23}^{+36}\pm (12.4)_{\langle{Z}\rangle}$  \\
\hline
$2.56-2.84$ & $-15.35$ & $-0.04039 \pm 0.02$ & $0.01949 \pm 0.023$ & $6.68 \pm (0.17)_{\text{stat}} \pm (1.15)_{\text{sys}}$ & $270$ & $263^{+286}_{-92} \pm (45)_{\langle{Z}\rangle}$ \\
\hline   
\end{tabular}
}
\endgroup
\caption{Moon shadow shifts in RA \& DEC with statistical errors, Moon shadow significance, and the calibrated median energy $E_{\textrm{cal}}$ (per $\log_{10}(N_{e\mu})$ bin), together with the composition/interaction model–averaged value of $\langle{Z}\rangle$ and the RT median energy $\langle \tilde{E}\rangle$.}
\label{tab:moon_shadow}
\end{table*}

Using the model-averaged values of $\langle {Z}\rangle$ and shifts along RA, we calculate the calibrated energy as follows.
\begin{equation}
E_{\textrm{cal}}=-1.59^\circ\langle {Z}\rangle /\Delta_{\textrm{data}} (\text{RA}).
\label{eq:cali}
\end{equation}
Here $\langle {Z}\rangle$, is averaged over four composition models and two interaction models. The values of $\langle{Z}\rangle$ along with the statistical and systematic errors are reported in the fifth column of \autoref{tab:moon_shadow}. The systematic error is defined by the standard deviation of the values of $\langle{Z}\rangle$ for the selected composition and interaction models. Both of these uncertainties (stat and sys) in $\langle {Z}\rangle$ are combined quadratically and propagated to $E_{\text{cal}}$ along with the statistical error in $\Delta_{\textrm{data}} (\text{RA})$  through \autoref{eq:cali}, affecting the final result of $E_{\text{cal}}$. The resultant values of $E_{\text{cal}}$ are reported in the last column of \autoref{tab:moon_shadow}. The asymmetric statistical error in $E_{\text{cal}}$ due to $\Delta_{\textrm{data}} (\text{RA})$ dominates over the error due to  $\langle{Z}\rangle$. 

\begin{figure}[H]
    \centering
\includegraphics[width=0.47\textwidth]{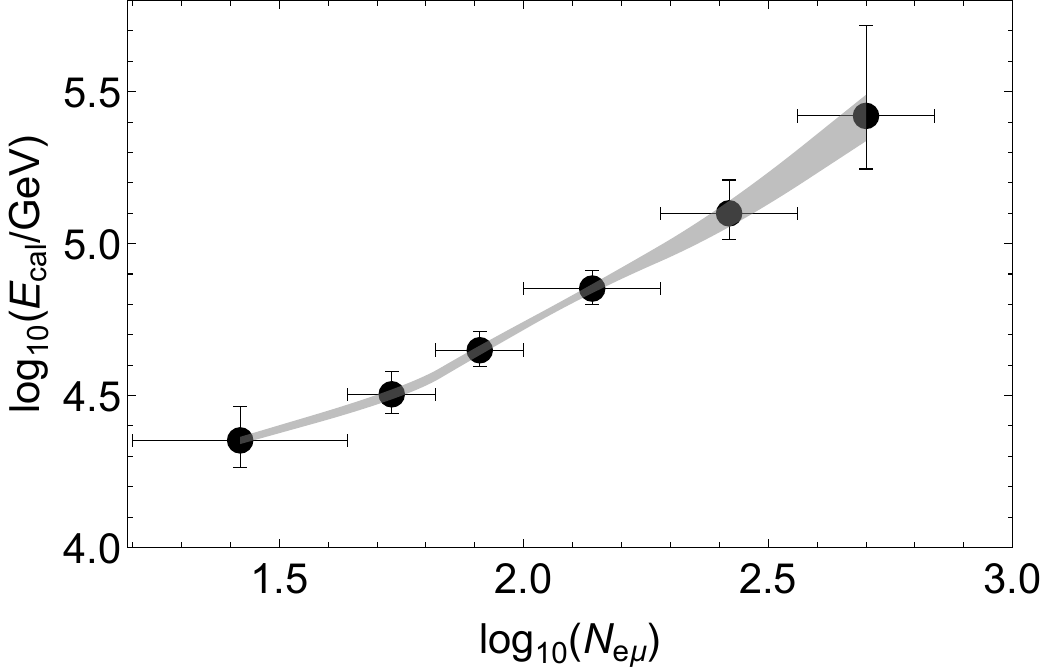}   
    \caption{Energy calibration is performed using the measured data. Black dots represent the data points, with error bars denoting statistical uncertainties ($\Delta_{\textrm{data}}$), while the gray band represents the systematic uncertainty obtained from the quadrature of the RT term and the composition/interaction model term (see \autoref{tab:moon_shadow}).
}
\label{ppr/fig:energy calibration}
\end{figure}

\begin{figure}[ht]
    \centering
\includegraphics[width=0.47\textwidth]{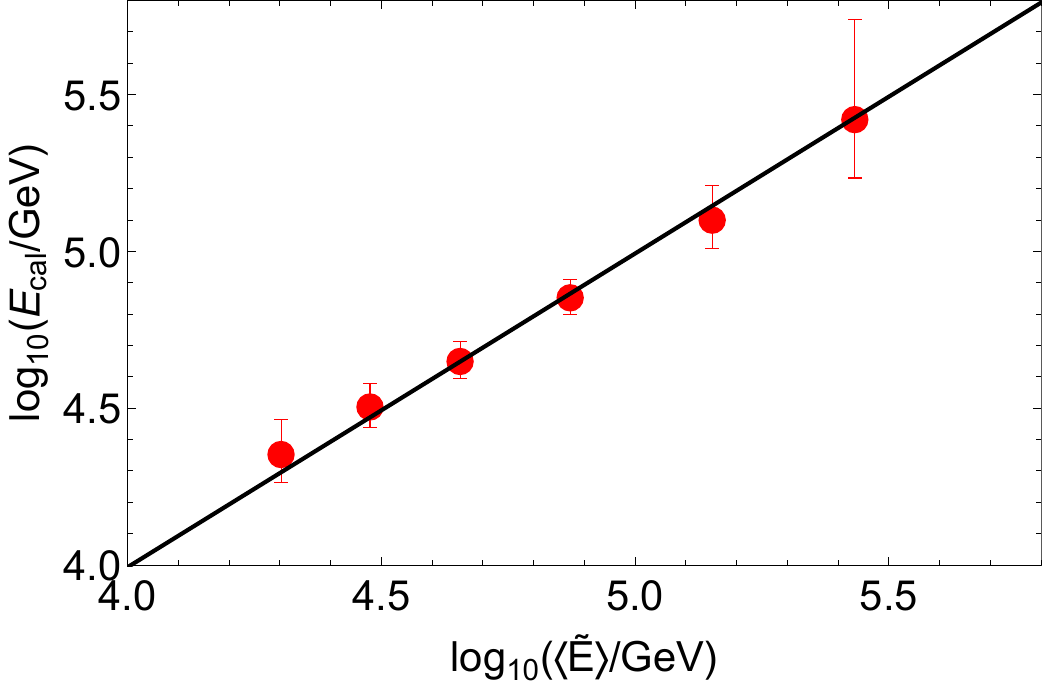}   
    \caption{Validation of the energy estimation across the calibrated range. The red data points show the composition-averaged median energy $\langle \tilde{E}\rangle$ versus the calibrated energy ($E_{\mathrm{cal}}$), with statistical uncertainties. A $\chi^{2}$ fit (black line) yields a best-fit parameter of $\epsilon = 0.015 \pm 0.08$.}
\label{ppr/fig:RecEvsTrueE}
\end{figure}

In \autoref{ppr/fig:energy calibration}, we plot \( E_{\text{cal}} \) the measured data for calibration (black points). The error bars in the data correspond to the statistical error in \( E_{\text{cal}} \) due to $\Delta_{\textrm{data}} (\text{RA})$, whereas the gray shaded region represents the error due to \( \langle{Z} \rangle \). 

In order to quantify the difference between calibrated and model-averaged median energies, we relate them by a scaling factor $(1-\epsilon)$ as follows:
\begin{equation}
E_{\textrm{cal}}=(1-\epsilon)\langle \tilde{E}\rangle.
\label{eq:epsilon}
\end{equation}

A chi-squared ($\chi^{2}$) minimization was performed to determine the best-fit value of the scaling parameter $\epsilon = 0.015 \pm 0.08 $, This implies that 
the energy scale is estimated with a 95\% confidence interval (CI) of $[-14\%,\, +17\%]$. 

In \autoref{ppr/fig:RecEvsTrueE}, the calibrated energies are fitted against the composition-averaged median energies $\langle \tilde{E}\rangle$ obtained from MC  via the scaling relation defined in \autoref{eq:epsilon}. The excellent agreement, characterized by only 8\% bias of $\epsilon$, validates our KM2A-MC simulations and confirms its accuracy within the calibrated range. This successful validation stems from our robust calibration procedure shown in \autoref{ppr/fig:energy calibration},  whose systematic uncertainties are modest across the first four bins and then increase toward higher energies. Together, these empirical verifications provide strong confidence in using KM2A-MC for reliable energy estimation at higher energies, with the fit re-performed in the high energy window, where direct calibration is not feasible.

\subsection{Uncertainties in energy calibration}

The systematic uncertainties in the calibrated energies, through the cosmic ray Moon shadow measurement, arise from three factors: the geomagnetic field model, the composition models, and the high energy hadronic interaction model. Comparison of direct measurements of the magnetic field near the surface of Earth with the T-IGRF model shows a difference of less than 1\% \cite{zhamin,2011igrf}. Considering the pointing accuracy of LHAASO-KM2A, the uncertainty due to GMF is negligible. The errors due to composition and interaction models propagate to the calibrated energy through the model-averaged value of $\langle{Z}\rangle $. As discussed earlier, we have calculated $\langle{Z}\rangle$ for eight different possible choices of composition and interaction models. The systematic error in $\langle{Z}\rangle$ is therefore defined by the standard deviation of these values of $\langle{Z}\rangle$ (see \autoref{tab:moon_shadow}). The calculated values of $\langle{Z}\rangle$ also carry a small statistical error of the MC simulation used in the ray-tracing method. 
The error due to pointing is small $0.009^\circ$ compared to others, bin-wise it remains below 8\% in the 20 to 70 TeV range, and increases to about 12\% or 22\% in the last two bins, corresponding to 125 and 260 TeV, respectively. However, the dominant error in the calibrated energies is statistical, resulting from the measured values of $\Delta_{\textrm{data}}(\text{RA})$ shown in the third column of \autoref{tab:moon_shadow}.

\subsection{Stability of the Detector}

\begin{figure}[H]
\centering
\includegraphics[width=0.45\textwidth]{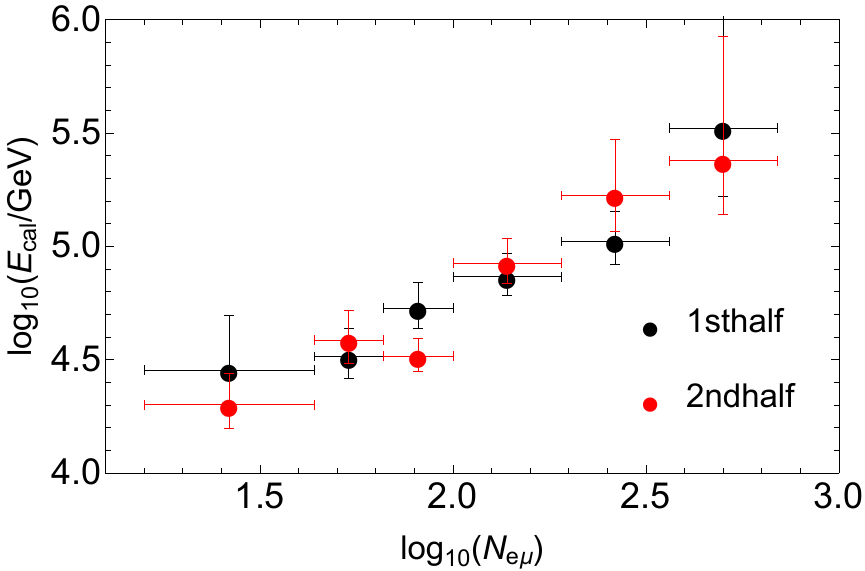}  
    \caption{The plot represents the stability of the detector over the long period of 3.5 years, dividing into two halves. The plot shows a comparison of calibrated energies.}
    \label{stability}
\end{figure}

\begin{figure}[H]
\centering\includegraphics[width=0.45\textwidth]{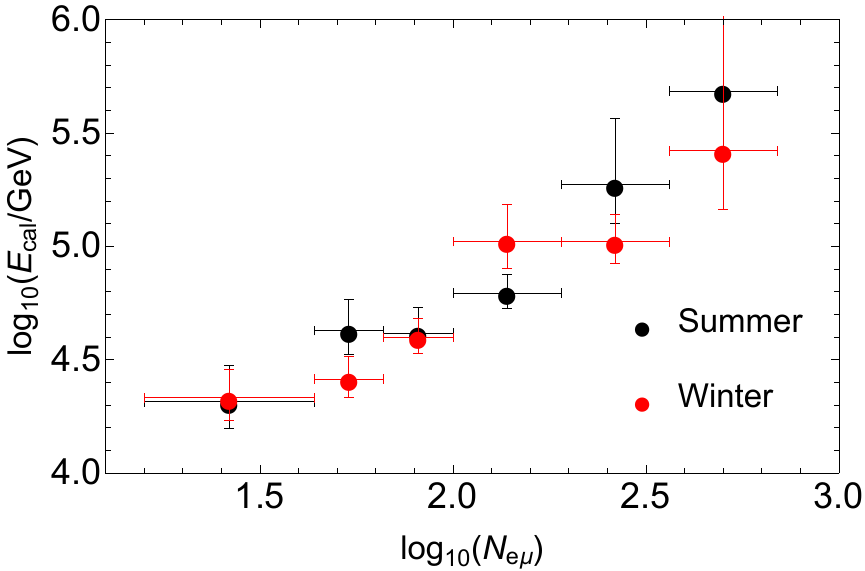} 
 \caption{To assess seasonal stability, we partitioned the  data set into two seasonal subsets: Summers (March–August; 18.5 months total) and Winters (September–February; 23 months). Seasonal segmentation shows no significant effect on the energy calibration.}
    \label{season}
\end{figure}

\autoref{stability} shows that the LHAASO–KM2A detector response is stable over 42 months (21 July 2021–31 January 2025). This duration includes seasonal variations such as the rainy season and fluctuating weather conditions at the LHAASO site. We divided the data into two equal time halves; their calibrated energies agree within uncertainties across all bins. Larger errors and small offsets in the lowest bins are consistent with reduced efficiency below 60 TeV, while the highest bin is statistics-limited. Overall, the calibrated energies fluctuate within the expected statistical and systematic uncertainties, with no evidence of drift over 3.5 years. To assess seasonal effects, we further partitioned the data into Summers (March–August; 18.5 months) and Winters (September–February; 23 months) only, and evaluated each subset separately. Results are reported in \autoref{season}; despite extreme weather at the LHAASO site, no significant seasonal effect is observed, and the long-term stability supports subsequent LHAASO–KM2A analyses, including Moon shadow studies and precise energy-scale calibration.

\newpage

\section{Conclusion}
\label{sec:conclusion}
In this study, we present results on the calibration of the LHAASO-KM2A detector using the Moon shadow technique in the energy range 20 to 260 TeV. The Moon shadow has been detected with high significance up to 48$\sigma$. The angular resolution, determined through 1D projection of deficit counts of events along the RA/DEC, varies from $0.41\pm0.03$ to $0.18\pm0.02$ degree in the given energy range, without accounting for the Moon's disc. Additionally, the detector's pointing accuracy is established as \( 0.0002^\circ \pm 0.009^\circ \), confirming the precision of the array's alignment. We have used the energy estimator $(N_{e\mu})$, whose response is nearly composition-independent across our calibrated range. Detailed studies of the method's bias and resolution are reported in \cite{zhyprl}.
The dominant uncertainties in the calibrated values of energies are due to the statistical error in the measured values of shifts along the RA direction. This error can only be reduced by acquiring more data.
Systematic uncertainties due to composition and interaction models are significantly less as compared to the statistical error in calibrated energies, 3\% in the energy range of 20 to 70 TeV, 9\% and 17\% at 125 and 260 TeV, respectively. 
Furthermore, we find that the overall difference between the calibrated values of energies obtained through Moon's shadow and the corresponding median energies obtained through KM2A-MC is about 8\% with a 95\% confidence interval of $[-14\%,\, +17\%]$. This result establishes the accuracy of the KM2A-MC in simulating the response of the detector, particularly within the calibrated range.

\section{ACKNOWLEDGMENTS}
We want to thank all staff members who work at the LHAASO site 4400 meters above sea level year-round to maintain a stable detector and other experiment components operating smoothly. We are grateful to Chengdu Management Committee of Tianfu New Area for the constant financial support for research with LHAASO data. We appreciate the computing and data service support provided by the National High Energy Physics Data Center for the data analysis in this paper. The following grants support this research work: The National Key R\&D program of China  2024YFA1611401, 2024YFA1611402, 2024YFA1611403, 2024YFA1611404, CAS-TWAS funding, the National Natural Science Foundation of China NSFC No.12275280, No.12393851, No.12393852, No.12393853, No.12393854, No.12205314, No.12105301, No.12305120, No.12105294, No.U1931201, No.12375107,  No.12261160362 and in Thailand by the National Science and Technology Development Agency (NSTDA) and the National Research Council of Thailand (NRCT) under the High-Potential Research Team Grant Program (N42A650868).
\newpage
\bibliographystyle{apsrev4-2}
\bibliography{References_PRD}
\end{document}